\documentclass{article}
\usepackage{spconf,amsmath,graphicx,hyperref}

\usepackage{cite}
\usepackage{amsmath,amssymb,amsfonts}
\usepackage{algorithmic}
\usepackage{graphicx}
\usepackage{textcomp}
\usepackage{xcolor}

\usepackage{bm, amssymb, booktabs, multirow,cite, url, xcolor, hyperref}
\usepackage{threeparttable}

\usepackage{etoolbox}
\apptocmd{\thebibliography}{\setlength{\itemsep}{0.25ex}\setlength{\parskip}{0pt}}{}{}

\title{Spatially Aware Self-Supervised Models for Multi-Channel Neural Speaker Diarization}
\name{{\shortstack[c]{
      Jiangyu Han$^{1}$,
      Ruoyu Wang$^{2}$,
      Yoshiki Masuyama$^{3}$,
      Marc Delcroix$^{4}$,\\
      Johan Rohdin$^{1}$,
      Jun Du$^{2}$, 
      Lukáš Burget$^{1}$
}}}
\address{$^{1}$Brno University of Technology, Czechia \;
$^{2}$University of Science and Technology of China, China \; \\
$^{3}$Mitsubishi Electric Research Laboratories (MERL), USA \;
$^{4}$NTT, Inc., Japan}

\begin{document}
\ninept
\maketitle
\begin{abstract}
Self-supervised models such as WavLM have demonstrated strong performance for neural speaker diarization. However, these models are typically pre-trained on single-channel recordings, limiting their effectiveness in multi-channel scenarios. Existing diarization systems built on these models often rely on DOVER-Lap to combine outputs from individual channels. Although effective, this approach incurs substantial computational overhead and fails to fully exploit spatial information. In this work, building on DiariZen, a pipeline that combines WavLM-based local end-to-end neural diarization with speaker embedding clustering, we introduce a lightweight approach to make pre-trained WavLM spatially aware by inserting channel communication modules into the early layers. Our method is agnostic to both the number of microphone channels and array topologies, ensuring broad applicability. We further propose to fuse multi-channel speaker embeddings by leveraging spatial attention weights. Evaluations on five public datasets show consistent improvements over single-channel baselines and demonstrate superior performance and efficiency compared with DOVER-Lap. Our source code is publicly available at \url{https://github.com/BUTSpeechFIT/DiariZen}.

\end{abstract}
\begin{keywords}
Multi-channel speaker diarization, DiariZen, self-supervised,  WavLM, cross-channel communication
\end{keywords}
\section{Introduction}
\label{sec:intro}
End-to-end neural diarization with vector clustering (EEND-VC) \cite{kinoshita2021integrating, kinoshita2021advances} has emerged as a key approach to speaker diarization. It integrates local EEND with speaker embedding clustering in a unified framework, enabling scalability to long conversations with many speakers. Recent advances in the local EEND module have been driven by self-supervised models such as WavLM \cite{chen2022wavlm, tawara2024ntt, han2025leveraging}. However, these models are typically pre-trained on single-channel audio, which limits their ability to exploit spatial information in multi-channel recordings.

Microphone arrays are increasingly common in meeting applications, where multi-channel spatial information helps distinguish speakers by their relative positions. To exploit such cues, prior studies have incorporated explicit multi-channel features \cite{pardo2006speaker, araki2008doa, hager2008handling, zheng2022multi, cord2025spatio} or applied attention mechanisms across channels \cite{horiguchi2022multi, wang2022cross, wu2023semi}, yet none have utilized self-supervised models. Another common approach is to run single-channel diarization independently on each microphone and then fuse the outputs with DOVER-Lap \cite{raj2021dover, kamo2026microphone, wang2025three}. While effective, methods based on DOVER-Lap are computationally demanding and fail to fully exploit spatial information. Moreover, most existing studies have been evaluated on limited benchmarks, raising concerns about generalization to diverse microphone array configurations.

In this work, we extend DiariZen \cite{han2025leveraging}, a speaker diarization pipeline built on EEND-VC that leverages WavLM \cite{chen2022wavlm} and Conformer \cite{gulati2020conformer} to strengthen the local EEND module. In the local EEND stage, we adapt the pre-trained single-channel WavLM to multi-channel diarization by introducing learnable channel communication modules into its early layers to capture spatial information. Our framework supports various channel communication mechanisms such as ChannelAttention (ChAtt) \cite{horiguchi2022multi, wang2022cross} and Transform-Average-Concatenate (TAC) \cite{luo2020end}, and is agnostic to both the number of channels and the microphone array topology. To further improve efficiency, we incorporate pruned WavLM \cite{han2025fine, han2025efficient}, which retains strong diarization performance while significantly lowering model complexity. Finally, in the clustering stage, we propose a spatial-attention fusion strategy that effectively integrates multi-channel speaker embeddings, thereby improving clustering performance.

We comprehensively evaluate our method on five public datasets: AMI \cite{carletta2005ami, kraaij2005ami}, AISHELL-4 \cite{fu2021aishell}, AliMeeting \cite{yu2022m2met}, NOTSOFAR-1 \cite{vinnikov2024notsofar}, and CHiME-6 \cite{watanabe2020chime}. 
The results show consistent benefits over the single-channel baseline and demonstrate superior performance and efficiency compared with the channel fusion using DOVER-Lap. In addition, our EEND model based on pruned WavLM achieves performance comparable to the unpruned model while substantially reducing computational overhead.
Moreover, fusing speaker embeddings using multi-channel spatial attention weights in the clustering stage yields significant improvements on CHiME-6, approaching the state of the art \cite{wang2023ustc} with a much simpler and more efficient system that requires no speech separation modules.

\section{Multi-channel extensions}\label{sec:methods}
In this section, we extend our DiariZen \cite{han2025leveraging} speaker diarization system, derived from the pyannote pipeline \cite{bredin2023pyannote, plaquet2023powerset}. The system follows a two-stage EEND-VC approach: in the first stage, an EEND model detects speaker activities within local overlapping windows; in the second stage, speaker embeddings are extracted from the speech of each detected speaker in each window, and these embeddings are then clustered to map local speaker identities to global ones.

We first introduce the EEND model used in the DiariZen system and explain how we extend it for multi-channel processing. In Section~\ref{subsec:att_spk_emb}, we then describe how multi-channel input can be leveraged to obtain more informative speaker embeddings for the second stage.

\subsection{Multi-channel EEND}

\begin{figure}[tbp]
  \centering
    \includegraphics[width=7.8cm]{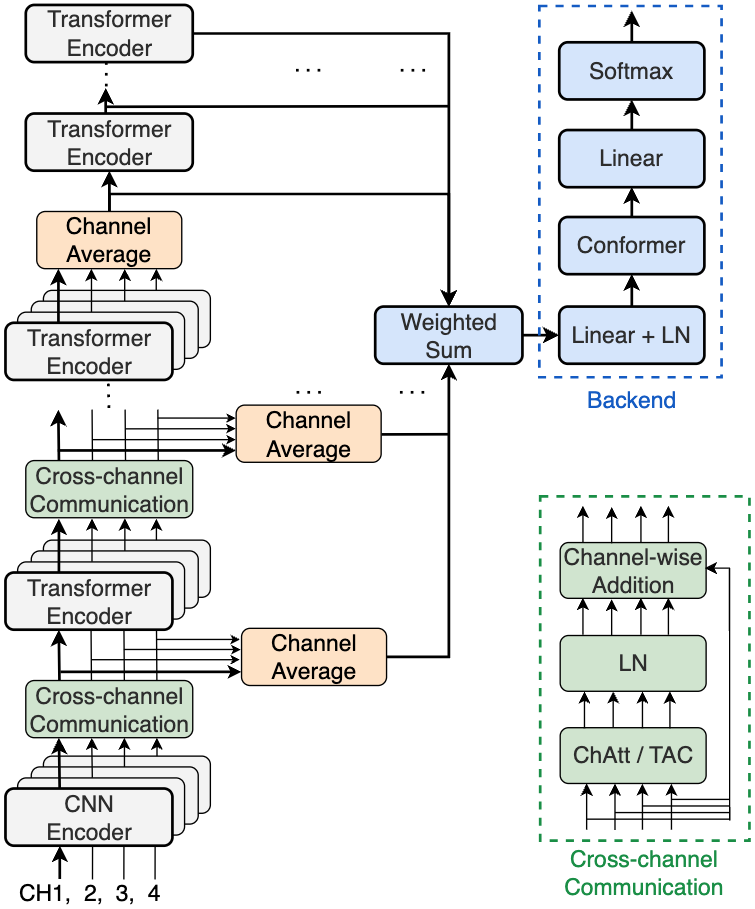}
  \caption{ 
    Framework of the multi-channel WavLM extension with four input microphones and its application to speaker diarization.}
  \label{fig:framework}
  \vspace{-0.5cm}
\end{figure}

The multi-channel extension of the EEND model in the DiariZen system is illustrated in Figure~\ref{fig:framework}, where LN denotes LayerNorm. Considering only the path from the first-channel input (bold arrows) and the processing blocks with bold outlines (ignoring the {\em Channel Average} and {\em Cross-channel Communication} blocks) recovers the original single-channel EEND model on which the extension is based. This model uses pre-trained WavLM~\cite{chen2022wavlm} (gray {\em Encoder} blocks) as a front-end. The output sequences of all WavLM layers are combined via a SUPERB-style weighted sum~\cite{yang2021superb} into a single sequence, which is fed into a Conformer-based backend~\cite{gulati2020conformer} predicting speaker activities as powerset classes~\cite{plaquet2023powerset}. Further details on the single-channel model and its training can be found in \cite{han2025leveraging}.

Following the work on multi-channel speaker verification~\cite{movsner2024multi}, we extend EEND as follows. Several early WavLM layers (stacked gray blocks in Figure~\ref{fig:framework}) are run in parallel on each channel (with parameters shared).  After each such layer, the outputs are averaged across channels ({\em Channel Average} blocks, in pale orange). These per-layer averaged sequences are then used in the same SUPERB-style combination as in the single-channel case to provide the input to the back-end. To reduce computation, the later WavLM layers operate directly on the aggregated input in a single-channel manner.

To exploit spatial cues encoded in the multi-channel input, we add trainable {\em Cross-channel Communication} blocks after each multi-channel encoder layer. We examine two variants of such blocks, both invariant to the number of channels and the microphone array configuration. In the first variant, ChannelAttention (ChAtt)~\cite{horiguchi2022multi, wang2022cross}, information is exchanged across channels using a standard multi-head self-attention \cite{vaswani2017attention} layer, followed by LayerNorm and a residual connection. 
% This is identical to the processing in WavLM {\em Transformer Encoder} layers, except that attention is applied per-frame across channels rather than per-channel across frames. 
In the second variant, the multi-head self-attention is replaced by the Transform-Average-Concatenate (TAC) mechanism \cite{luo2020end}, while retaining LayerNorm and the residual connection. For each frame, TAC applies a shared linear layer with PReLU activation to the input from each channel; the transformed representations are then averaged across channels to obtain a global vector, which is concatenated with each original channel input and projected back to the original dimensionality by a linear layer.

To train the multi-channel EEND model, all parameters are initialized from the pre-trained single-channel model, except for those of the newly introduced Cross-channel Communication blocks. These blocks are initialized as identity mappings to preserve the behavior of the pre-trained model at the start of training. Specifically, the scale and bias parameters in the LayerNorm are set to zero, forcing the block output to zero and thereby passing its input directly to the output through the residual connection. This allows the subsequent fine-tuning on multi-channel data to begin from a stable single-channel baseline while gradually learning to exploit spatial information across microphone channels.

\vspace{-0.3cm}

\subsection{Spatially Attentive Speaker Embeddings}
\label{subsec:att_spk_emb}
The previous section described the multi-channel extension of the EEND model, used in the first stage of the DiariZen EEND-VC pipeline. Ideally, speaker embedding extraction in the second stage should also leverage multi-channel input—a topic of our ongoing research—but in this work we use a single-channel extractor and process each channel separately. In Section~\ref{subsec:VBx_results}, we analyze strategies to combine speaker embeddings across channels for optimal performance, including the two approaches proposed here.

These approaches exploit spatial attention weights from a pre-trained ChannelAttention module to select or combine embeddings. Let $\mathbf{S} \in \mathbb{R}^{T \times H \times C \times C}$ be the attention weights from a selected layer, with $T$ frames, $H$ heads, and $C$ channels. Averaging across frames and heads yields a global representation $\mathbf{S}_\mathrm{g} \in \mathbb{R}^{C \times C}$, as shown in Figure~\ref{fig:att_vis}. We then average $\mathbf{S}_\mathrm{g}$ over rows (queries) to obtain channel weights $\mathbf{\hat{S}}_\mathrm{g} \in \mathbb{R}^C$. The per-channel embeddings for each speaker in each local window are then combined across channels, either by (i) selecting the highest-weighted channel ({\em attentive argmax}) or (ii) computing a weighted average ({\em attentive weighted fusion}). This procedure requires no additional training, as it directly uses intermediate representations from the pre-trained ChannelAttention module.

\section{Experiments}
\label{sec:experiments}
% \vspace{-0.3cm}
\subsection{Datasets}
\label{subsec: datasets}
We train and evaluate our system on a compound dataset combining five public datasets: AMI \cite{carletta2005ami, kraaij2005ami}, AISHELL-4 (AIS-4) \cite{fu2021aishell}, AliMeeting (Ali) \cite{yu2022m2met}, NOTSOFAR-1 (NSF-1) \cite{vinnikov2024notsofar}, and CHiME-6 (CH-6) \cite{watanabe2020chime}. Since AISHELL-4 lacks a development set, 10\% of its training data from each room is randomly selected for validation and the rest for training. For CHiME-6, we apply WPE \cite{drude2018nara} and BeamformIt \cite{anguera2007acoustic} to each array, generating beamformed audio and reducing the number of channels from 24 to 6.

\subsection{Configurations}
\label{subsec: configurations}

Our EEND models are based on WavLM Base+ and its pruned variant with 80\% sparsity, both initialized from the pretrained DiariZen models \cite{han2025efficient}. We insert Cross-channel Communication modules into the first four layers of WavLM models, with input and hidden dimensions of 768 and 256, respectively, and employ 8 heads for the ChannelAttention mechanism.  For model training and inference, we use the same hyper parameters as in \cite{han2025leveraging}. VBx \cite{landini2022bayesian} is used as the clustering method to determine the speaker mappings between the EEND local windows.
We measure performance with diarization error rate (DER), applying a 0.25-second collar for CHiME-6 and none for other datasets. We also report the macro-averaged DER to reflect overall performance across datasets. Our source code and detailed configurations are publicly available\footnote{\url{https://github.com/BUTSpeechFIT/DiariZen}}.

\subsection{Results and Discussion}
\subsubsection{Performance under oracle clustering}
Because the proposed multi-channel EEND extensions affect only the local EEND output, we first evaluate them with oracle clustering. Specifically, instead of performing clustering, the speakers detected in the local windows are assigned global (ground-truth) speaker labels so as to minimize the overall DER. Results are shown in Table~\ref{tab:oracle_clustering}.

As a baseline, we use single-channel systems, where \textbf{audio from the first channel} is used for both training and evaluation. For ChannelAttention (ChAtt) and Transform-Average-Concatenate (TAC), we use the first four channels to simplify the experiments and accelerate validation, while still retaining spatial information from multiple, physically distinct microphones.
A checkmark denotes the use of the pruned WavLM Base+ model with 80\% sparsity \cite{han2025efficient}; otherwise, the unpruned model is applied. Our earlier work \cite{han2025fine, han2025efficient} showed that up to 80\% of WavLM parameters can be removed through structured pruning, yielding substantial speedups without degrading diarization accuracy. However, its effectiveness in multi-channel processing has not been studied prior to this work.

As shown in Table~\ref{tab:oracle_clustering}, the pruned WavLM performs slightly worse under the single-channel condition, but this degradation disappears in multi-channel scenarios. Although extending pre-trained models to multi-channel has typically been computationally demanding, our results suggest that the pruned model can be effectively applied, substantially reducing computational complexity. We therefore adopt the pruned WavLM for all subsequent experiments.

Across most of datasets, different cross-channel communication methods achieve comparable performance, all outperforming the single-channel baseline.
%CoAttention \cite{horiguchi2022multi} was also evaluated but offered no clear advantage.
Therefore, we adopted ChannelAttention for the remaining experiments due to its simplicity and intuitive mechanism for multi-channel communication.

\subsubsection{Performance under VBx clustering}
\label{subsec:VBx_results}
While previous experiments demonstrate the effectiveness of our multi-channel EEND extensions, they are restricted to oracle clustering with four channels. To evaluate performance under more realistic conditions, Table~\ref{tab:vbx_clustering} presents results using VBx \cite{landini2022bayesian} for speaker embedding clustering, where all microphone channels are utilized.

The first section of Table~\ref{tab:vbx_clustering} reports baseline results using a single-channel EEND model trained on recordings from \textbf{randomly selected microphone channels}. The first line shows results with only the first channel as input. The second line corresponds to a widely used multi-channel diarization strategy~\cite{kamo2026microphone, wang2025three}, in which both stages of the EEND-VC pipeline are applied separately to each channel, and the channel-wise diarization outputs are then combined with DOVER-Lap~\cite{raj2021dover}.
The third line (Average probs \& embs) corresponds to a system where EEND is applied separately to each channel, but the output probabilities are averaged across channels to form a fused output. This fused output is then used as the frame-level speaker activity reference for extracting embeddings from each channel, which are subsequently averaged across channels for clustering. This simple averaging outperforms DOVER-Lap on CHiME-6, exposing its limitations in some conditions.

\begin{table}[tbp]
\vspace{-0.2cm}
  \caption{Performance comparison under oracle clustering. Multi-channel systems use first 4 channels for training.}
  \label{tab:oracle_clustering}
 \setlength{\tabcolsep}{1.2mm}
 \resizebox{\linewidth}{!}{%
  \centering
  \begin{tabular}{@{}l | c c| c c c c c| c@{}}
    \toprule
        \multirow{2}{*}{System} & 
        \multicolumn{2}{c|}{WavLM} & \multicolumn{5}{c|}{DER (\%)} & \multirow{2}{*}{Macro} \\
        & Pruned & Params & AMI & AIS-4 & Ali & NSF-1 & CH-6 &  \\
        \midrule
        \multirow{2}{*}{\begin{minipage}{0.45in} Single-channel\end{minipage}} & - & 94.4M & 13.5 & 8.9 & 12.5 & 14.2 & 24.9 & 14.8 \\
        & \checkmark & 18.8M & 13.3 & 9.3 & 12.7 & 14.2 & 25.7 & 15.0 \\
    \midrule
    ChAtt & - & 94.4M & 13.1 & 9.2 & 12.2 & 14.0 & 22.8 & 14.3 \\
     ChAtt & \checkmark & 18.8M & 12.9 & 9.0 & 11.8 & 13.8 & 22.9 & 14.1 \\
    TAC & \checkmark & 18.8M & 12.8 & 8.9 & 12.0 & 14.1 & 22.9 & 14.1 \\
    \bottomrule
  \end{tabular}}
  \vspace{-0.3cm}
\end{table}

\begin{table}[tbp]
\vspace{-0.1cm}
  \caption{Performance comparison under VBx clustering. Multi-channel systems use all available channels for training.}
  \label{tab:vbx_clustering}
 \setlength{\tabcolsep}{1.2mm}
  \centering
  \resizebox{\linewidth}{!}{%
  \begin{tabular}{@{}l | c c c c c| c@{}}
    \toprule
        \multirow{2}{*}{System} & 
        \multicolumn{5}{c|}{DER (\%)} & \multirow{2}{*}{Macro} \\
        & AMI & AIS-4 & Ali & NSF-1 & CH-6 &  \\
        \midrule
        Single-channel  & 15.3 & 11.3 & 15.0 & 17.7 & 33.8 & 18.6 \\
        DOVER-Lap  & 14.7 & 10.9 & 13.5 & 17.1 & 30.9 & 17.4 \\
        Average probs \& embs & 14.9 & 11.0 & 14.0 & 17.5 & 28.8 & 17.2 \\
    \midrule
    ChAtt, DOVER-Lap & 14.8 & 11.0 & 12.8 & 17.4 & 31.3 & 17.5 \\
    ChAtt, average embed. & 14.9 & 11.1 & 12.9 & 17.6 & 28.5 & 17.0 \\
    ChAtt, att. argmax & 14.9 & 11.0 & 12.8 & 17.5 & 29.5 & 17.2 \\
    ChAtt, att. weighted fusion & 14.8 & 11.2 & 12.8 & 17.4 & 27.5 & 16.7 \\
    \bottomrule
  \end{tabular}}
  \vspace{-0.3cm}
\end{table}

The following results are obtained with the multi-channel EEND model using ChannelAttention, which produces a single diarization output. However, speaker embeddings are still extracted separately from each channel. For the line “ChAtt, DOVER-Lap”, clustering is applied independently to each channel, yielding per-channel global diarization outputs that are then combined with DOVER-Lap. This approach clearly outperforms the single-channel baseline, with large gains on AliMeeting and CHiME-6, but offers no extra advantage over the other two baselines except on AliMeeting.
In the next line, speaker embeddings are averaged across channels before clustering, yielding a single global diarization output. This reduces the DER on CHiME-6 to 28.5, demonstrating the benefit of incorporating spatial information into the embeddings through simple aggregation.

The last two lines of Table~\ref{tab:vbx_clustering} show results for two multi-channel approaches introduced in Section~\ref{subsec:att_spk_emb}: {\em attentive argmax} for selecting embeddings, and {\em attentive weighted fusion} for combining embeddings.
Relative to the DOVER-Lap baseline in the second row, the attentive argmax approach shows substantial gains on AliMeeting and CHiME-6 while performing comparably on the other datasets. It is also more efficient, since local EEND and speaker embedding clustering are performed only once rather than per channel. On CHiME-6, attentive weighted fusion further reduces DER to 27.5, approaching the performance of top-ranked systems \cite{wang2023ustc}, while maintaining a simpler design without speech separation.

\subsubsection{Analysis of inference time}
\begin{figure}[tbp]
  \centering
  \includegraphics[width=8.2cm]{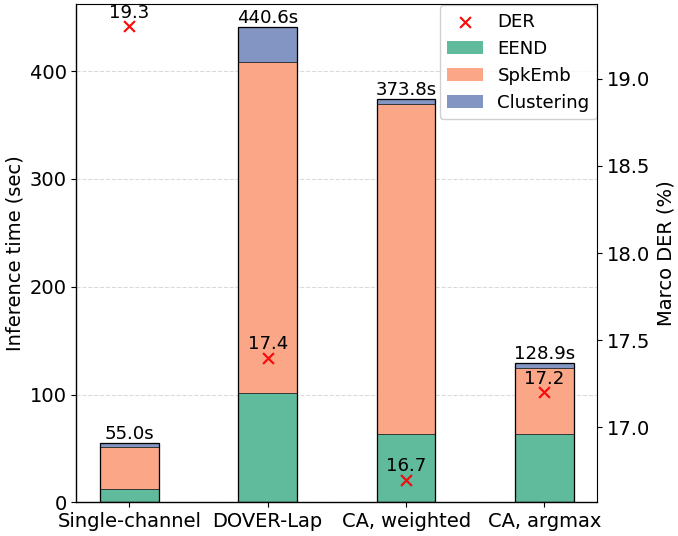}
  \caption{ 
    Inference time at different stages for various methods. The results are averaged over five runs. CA denotes ChannelAttention.}
  \label{fig:time_compare}
  \vspace{-0.3cm}
\end{figure}

In Figure~\ref{fig:time_compare}, we compare the inference time of different methods across stages on a single A5000 GPU. Inference uses a batch size of 32 and the AMI EN2002a recording, about 2143 seconds long with eight channels. The macro DER over all datasets is also reported.
Compared with the single-channel baseline, DOVER-Lap substantially improves performance but incurs considerable computational overhead. Our method is consistently more efficient than DOVER-Lap, particularly with the attentive argmax approach. Among the three stages, speaker-embedding extraction dominates computational cost. Extracting embeddings only from the microphone channel with the highest attention score can significantly reduce this overhead. These findings indicate that our method provides a more practical and efficient alternative to DOVER-Lap.

\subsubsection{Analysis of attention scores}
\label{subsubsec: att_score}
\begin{figure}[tbp]
  \centering
  \includegraphics[width=8.6cm]{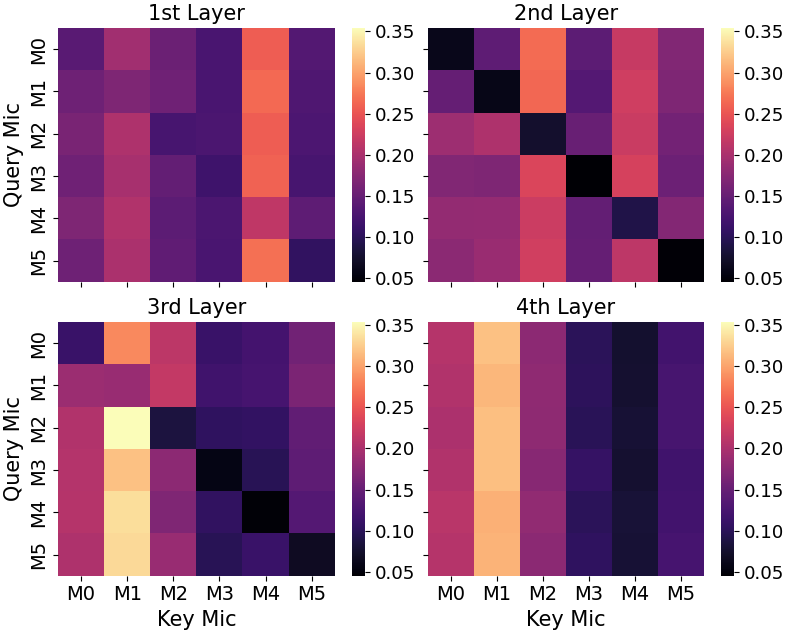}
  \caption{ 
    Layer-wise channel attention weights averaged over frames and attention heads on CHiME-6 (S21, 100th chunk).}
  \label{fig:att_vis}
  \vspace{-0.3cm}
\end{figure}

Since our method achieves the largest gains on CHiME-6, we further analyze the channel attention weights using the 100th chunk of session S21 from the CHiME-6 evaluation set. As shown in Figure~\ref{fig:att_vis}, results indicate that different layers exhibit distinct behaviors: in the second layer, the model tends to ignore the current microphone, whereas in the fourth layer all queries assign higher weights to the first three channels. Moreover, we observe that attention patterns vary across chunks and sessions, indicating that the system adapts to speaker-position changes by exploiting multi-channel spatial cues.

\begin{figure}[tbp]
  \centering
  \includegraphics[width=8.6cm]{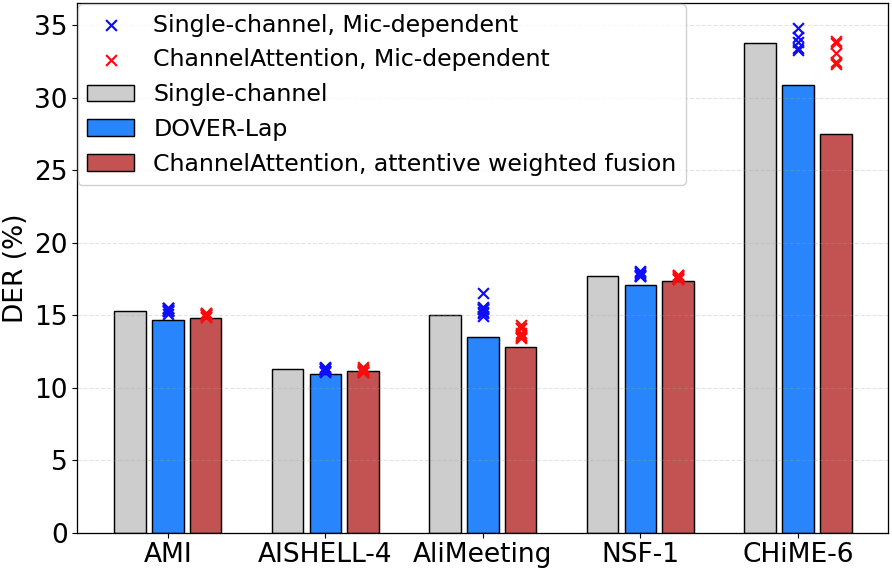}
  \caption{ 
    Performance comparison of single-channel (Mic-dependent) and multi-channel methods. NSF-1 denotes NOTSOFAR-1 dataset.}
  \label{fig:mic_depend}
  \vspace{-0.3cm}
\end{figure}

\subsubsection{Analysis of microphone dependence}
Our earlier experiments demonstrate substantial multi-channel improvements on AliMeeting and CHiME-6, while gains on the other three datasets are more modest. To understand this, we examine two systems: a single-channel model trained on randomly selected microphone audio, and ChannelAttention trained on all available microphones. At inference, the single-channel model is applied separately to each microphone channel to obtain microphone-dependent (Mic-dependent) results. For ChannelAttention, similar Mic-dependent results are derived by clustering speaker embeddings extracted from each microphone individually.

Figure~\ref{fig:mic_depend} compares the Mic-dependent results across all channels (red and blue cross  markers, $\times$) with the single-channel system, DOVER-Lap fusion, and ChannelAttention with attentive weighted fusion. Substantial variance in Mic-dependent DERs is observed for AliMeeting and CHiME-6, likely due to their recording configurations. For instance, CHiME-6 microphones are distributed across multiple rooms, leading to pronounced performance gaps. This variability highlights the value of spatial information and explains the observed gains. By comparison, AMI, AISHELL-4, and NOTSOFAR-1 exhibit more consistent recording conditions, which limits the potential benefits of multi-channel modeling.

\section{Conclusion}
\label{sec:conclusion}
In this work, we proposed an efficient and general approach for extending pre-trained single-channel EEND models to multi-channel speaker diarization. By inserting channel communication modules into the early layers and initializing them to preserve the pre-trained model’s behavior, our method captures spatial cues without disrupting original representations. We further introduced strategies to combine speaker embeddings using spatial attention weights, requiring no additional training. Importantly, the approach is agnostic to the number of microphone channels and array topology.
Experiments on five public datasets show that our method consistently outperforms the single-channel baseline and surpasses DOVER-Lap, while being significantly more efficient. On CHiME-6, it achieves strong performance without relying on speech separation. These results highlight the practicality and scalability of the proposed framework for real-world multi-channel diarization.

% \newpage
\bibliographystyle{IEEEbib}
\bibliography{mybib}

\end{document}